\documentclass[prd,showpacs,altaffilletter,twocolumn,
superscriptaddress,floatfix,nofootinbib,preprintnumbers]{revtex4-1} 

\pdfoutput=1
\usepackage{enumerate} 
\usepackage{amsfonts,amsmath,graphicx,bm}
\usepackage{dcolumn}
\usepackage[pdftex,
       colorlinks=true,
       urlcolor=blue,       
       filecolor=green,     
       linkcolor=red,       
       pdftitle={Exploring S-Wave Threshold Effects in QCD: A Heavy-Light Approach},
       pdfpagemode=None,
       bookmarksopen=true]{hyperref}

\usepackage{ifpdf}
\usepackage{multirow}
\usepackage[colorlinks=true]{hyperref}
\usepackage{url}
\usepackage[caption=false]{subfig}
\usepackage[utf8]{inputenc}

\newcommand{\fermilab}{Fermi National Accelerator Laboratory, Batavia, Illinois, 60510, USA}

\begin{document}
\title{Exploring S-Wave Threshold Effects in QCD: A Heavy-Light Approach}

\author{Estia \surname{Eichten}}
\email[]{eichten@fnal.gov}
\affiliation{\fermilab}

\author{Ciaran \surname{Hughes}} 
\email[]{chughes@fnal.gov}
\affiliation{\fermilab}

\pacs{12.38.Gc, 13.20.Gd, 13.40.Hq, 14.40.Pq}
\preprint{FERMILAB-PUB-19-558-T}

\begin{abstract}
  QCD exhibits complex dynamics near S-wave two-body thresholds.  
  For light mesons,  we see this in the failure of quark models to explain the $f_0(500)$ and $K_0^*(700)$ masses. 
  For charmonium, an unexpected $X(3872)$ state appears at the open charm threshold. 
  In heavy-light systems, analogous threshold effects appear for the lowest $J^P = 0^+$ and $1^+$ states in the $D_s$ and $B_s$  systems.
  Here we describe how lattice QCD can be used to understand these threshold dynamics by smoothly varying the strange-quark mass when studying the heavy-light systems. Small perturbations around the physical strange quark mass are used so to always remain near the physical QCD dynamics. This calculation is a straightforward extension of those already in the literature and can be undertaken by multiple lattice QCD collaborations with minimal computational cost. 
\end{abstract}

\maketitle

\section{Introduction}
\label{sec:intro}

The dynamics of QCD simplifies for quark masses in two limits. First, for light-quarks $q =  u, d$ and $s$ whose quark masses are small compared to the confining scale $\Lambda_{QCD}$,  spontaneously broken chiral symmetry can be used to predict the masses and interactions of low-lying pseudoscalar mesons. Secondly, for heavy quarks $Q= c, b$ and $t$, whose quark masses are large compared to $\Lambda_{QCD}$, the resulting separation of physical scales yields both a qualitative and surprisingly accurate quantitative phenomenological understanding of this sector.

However, phenomenological calculations in these sectors have certain limitations. In the first scenario above, early results using quark models with phenomenological potentials were generally useful guides to understand the properties of mesons and baryons involving only light quarks. Yet they fail to incorporate the effects of QCD chiral symmetry.  Particularly striking is the failure to understand the nature of ground states in the  $J^{P} = 0^{+}$ channels (e.g. the $\sigma [f_0(500)]$ and $\kappa [K_0^*(700)]$), which are seen in the analysis of S-wave  $\pi \pi$ and $K \pi$ scattering, but not expected in quark models. Today these models have been superseded by direct Lattice QCD (LQCD) calculations \cite{Wilson:2019wfr, Briceno:2016mjc}.

In the second scenario, for heavy-heavy mesons the heavy-quark velocity $p_Q/m_Q \approx v/c$ is small,  which allows for a nonrelativistic effective field theory description.  The 
gluons and light quark interactions are seen by the heavy  quarks as effective confining potentials, which can be computed by LQCD or modeled by phenomenological potentials. The low-lying spectrum  can then be calculated using the Schrodinger equation for the heavy quark system. For states below threshold, this produces robust  predictions for masses, decays and transitions (for Zweig allowed strong decays) \cite{Brambilla:2010cs}. However, above threshold the dynamics are more complicated.  For the charmonium system, a number of possible new states  (called the XYZ states) have been observed experimentally \cite{PhysRevD.98.030001}. The first of these states, the $X(3872)$, was observed in 2003 by Belle \cite{Choi:2003ue} and quickly confirmed by BaBAR \cite{Aubert:2004ns}, CDF \cite{Acosta:2003zx} and D0 \cite{Abazov:2004kp}.  It is a surprisingly narrow $J^{PC}=1^{++}$ state, very close to the S-wave $D^{0 *} \bar D^{0}$ threshold. 

Finally, for heavy-light mesons the heavy-quark can be viewed as a static source for the light degrees of freedom in leading order heavy quark effective theories\footnote{In fact, these systems are ideally suited for studying the interplay of chiral symmetry breaking and confinement in QCD as a single dynamical light (valance) quark is coupled to a static color source.}.  Corrections to this leading behavior can be found with expansion parameter $\Lambda_{\text{QCD}}/m_Q$.  Here too various relativistic quark models were used to calculate the light quark dynamics and the excitation spectrum of these mesons.
Again these models failed to predict \cite{Godfrey:1985xj,DiPierro:2001dwf} the narrow $D_{s0}^*(2317)$ and $D_{s1}(2460)$ states observed by BaBar \cite{BABAR:Ds0,BABAR:Ds1} and Belle \cite{BELLE:Ds0, BELLE:Ds1} in 2003, which are found slightly below the S-wave  $DK$ and $D^*K$ thresholds.
All these examples show that QCD dynamics near S-wave thresholds are strikingly strong and more complicated than expected.

Theorists have suggested a large number of models to explain the QCD effects around thresholds.  Some models propose new states arising from the strong interactions between the two mesons at 
a S-wave threshold.  Here the dynamical  pictures include tetra-quarks (compact states with two valence quarks and two valence anti-quarks in various configurations) \cite{Jaffe:2004ph} or molecular states 
(loosely-bound two-meson states) \cite{Du:2017zvv}.  Other models do not introduce new states, but argue that the properties of single meson states are greatly modified by mixing with two meson contributions \cite{Hwang:2004cd}. Others suggest that the residual effects of the confinement and chiral symmetry breaking interplay may be important \cite{Bardeen:2003kt}. 

More experimental data will help clarify and constrain models in systems where theorists can make reliable predictions.
However, the overlap between theoretically tractable systems which are also presently experimentally accessible is small.  The purpose of this work is to show how to efficiently employ lattice QCD calculations in heavy-light systems to resolve the theoretical situation. 

Notably, the parameters used in a lattice QCD calculation do not need to be fixed to the values found in nature. For example, it is possible to smoothly vary the quark masses and examine how physical systems change as a result. We utilize this to supplement the experimental data (which is limited to the physical values of quark masses). 
In fact, many lattice QCD calculations are already done at unphysical up, down, and strange quark mass, either to reduce the computational time or to aid in some form of interpolation/extrapolation. 
The Hadron Spectroscopy collaboration, for example, have studied the effect of having light-quark masses that produce $M_{\pi}=236$ MeV or $391$ MeV, and show that the $\sigma$ becomes stable for the latter case \cite{Briceno:2016mjc}.
Consequently, results far from the physical point may differ in substantive ways from the behavior of full QCD, particularly with regard to the interplay of chiral symmetry breaking and confinement.
In order to make progress, it is necessary to specify systems surrounding thresholds which are computationally cheap, theoretically simple, and allow a small smooth variation of the quark mass around the physical point to open or close the lowest threshold.  We show  how to address all these points using heavy-light meson systems, and focus on the $D_{s0}\to DK$ and $B_{s0}\to BK$ channels.

This paper is organized as follows. In Sec.~\ref{sec:HQET} we present a brief overview of the application of heavy quark effective theory to the $D_s$ and $B_s$ heavy-light systems. 
Sec.~\ref{sec:chi} briefly reviews spontaneously broken chiral symmetry, with application to pseudoscalar bosons such as the kaon. The main results of this work are found in Sec.~\ref{sec:simple}. There, we describe how the strange quark mass can be varied by small perturbations in order to make the $D_{s0}/B_{s0}$ mass lie on top of the $DK/BK$ threshold. In Sec.~\ref{sec:lattice} we discuss how studying the $D_{s0}/B_{s0}$ for various strange quark masses in a  lattice QCD calculation is a practical proposal, and is a straightforward extension of work already found in the literature. Finally, we summarize in Sec.~\ref{sec:conclusions}. 

\section{Heavy Quark Effective Field Theory and Heavy-Light Mesons}
\label{sec:HQET}

For a heavy-quark $Q$ with mass $m_Q \gg \Lambda_{\rm QCD}$, the HQET Lagrangian is given by
\begin{eqnarray}
{\cal L}_\psi  &=&   \psi^{\dagger}( i{\cal D} _0)
 \psi - \psi ^{\dagger} m_Q \psi  
        + {1\over 2m_Q} \psi ^{\dagger} \vec{{\cal D}}^2 \psi \nonumber \\
&& + c_F {g\over 2m_Q} \psi ^{\dagger} \vec{\sigma}\cdot 
   \vec{B} \psi + \mathcal{O}\left({\Lambda_{QCD}^2\over m_Q^2}\right) \;,
\label{HQET}
\end{eqnarray}
where  $\psi \equiv {(1+\gamma_0)} \psi _{\rm Dirac}$ is a two 
component quark field, $B^i \equiv {1\over 2}g t^a\epsilon ^{ijk} F_{jk}^a$,
and ${\cal D}_{\mu} = \partial _{\mu} - i g t^a A_{\mu}^a$.
The effective interactions of light quarks and gluons remain unchanged through order $\mathcal{O}({\Lambda_{QCD}/ m_Q})$.

To leading order  the heavy quark propagates only in time and provides a color source for the associated light system.  Thus, for hadrons with only one heavy quark, the dynamics of the system are independent of both the heavy quark mass $m_Q$ and spin $S_Q$.  In particular for heavy-light mesons in this limit, the  total angular momentum and parity of the light degrees 
of  freedom, $j_l^P$,  are good  quantum numbers, and each state is doubly degenerate associated with the two spins of the heavy quark. As such, the ground state has 
$j_l^P=\frac{1}{2}^-$ with total  $J^P = 0^-,1^-$. The first set of excited levels (the P states) are $j_l^P=\frac{1}{2}^+$ ($J^P=0^+, 1^+$) and $j_l^P=\frac{3}{2}^+$ ($J^P=1^+, 2^+$).  

Still, the $1/m_Q$ corrections shown in Eqn.~(\ref{HQET}) need to be considered. The spin dependent interactions will split the two-fold degeneracy in $S_Q$.  In addition there are spin independent $1/m_Q$ corrections which mix states differing by one unit in $j_l^P$ but with the same $J^P$. From the interactions in Eqn.~(\ref{HQET}), one can see that the heavy-light meson mass has a general dependence 
\begin{eqnarray}
M_{Q\bar q} &=& m_Q + m_q + {\cal C}_0(j_l^P, m_q) + {\cal C}_1( j_l^P, m_q)/m_Q  \nonumber \\
                    &+& (S\cdot j_l) {\cal C}_2(j_l^P, m_q)/m_Q + \mathcal{O}(\Lambda_{\rm QCD}^2/m_Q^2) \,. 
\label{eqn:Hlmass} 
\end{eqnarray}
The light quark and gluon dynamics are contained in the ${\cal C}_0$, ${\cal C}_1$ and ${\cal C}_2$ terms.
In order to more concisely see the light quark dependence of the ${\cal C}_i$ coefficients, we can include an arbitrary finite term in $m_Q$ by the redefinition $m_Q \to \tilde m_Q = m_Q + \tilde {\cal C}_0(\frac{1}{2}^-,0)$, where  $\tilde {\cal C}_0(j_l^P,m_q)= {\cal C}_0(j_l^P, m_q) + {\cal C}_1(j_l^P,m_q)/m_Q$. Then Eqn.~(\ref{eqn:Hlmass}) becomes 
\begin{eqnarray}
M_{Q\bar q} &=& \tilde m_Q + m_q + \tilde{\cal C}_0(j_l^P, m_q) - \tilde{\cal C}_0( \frac{1}{2}^-, 0)  \nonumber \\
                    &+& (S\cdot j^P_l) {\cal C}_2(j_l^P, m_q)/ m_Q + \mathcal{O}(\Lambda_{\rm QCD}^2/m_Q^2) \,.
\label{eqn:Hlredo} 
\end{eqnarray}
The $m_q$ dependence of $\tilde {\cal C}_i$ can be determined by computing its value for systems containing up/down quarks vs.~strange quarks. By taking the spin-average of $M_{Q\bar q}$ in a given $j_l^P$ multiplet, the $S\cdot j^P_l$ term in Eqn.~(\ref{eqn:Hlredo}) disappears.  Ignoring the very small $m_u$ dependence in ${\cal C}_1(\frac{1}{2}^-,m_u)$,  ${\cal C}_1(\frac{1}{2}^-,m_u) = {\cal C}_1(\frac{1}{2}^-,0)$  and thus for the $j_l^P=\frac{1}{2}^-$ $D_u$ and $B_u$ ground states the dependence on ${\cal C}_0$ also disappears from Eqn.~(\ref{eqn:Hlredo}). Thus we can define $\tilde m_c = (M_{D^0} + 3M_{D^{*0}})/4 - m_u$, and $\tilde m_b = (M_{B^-} + 3M_{B^{*-}})/4 - m_u$.  
Additionally, if we ignore any $m_q$ dependence in $\tilde{\cal C}_0(j_l^P, m_q)$,  the discrepancy between its determination for up/down quark vs.~strange quark systems will be caused by the small explicit $m_q$ dependence up to $\mathcal{O}(\Lambda_{\rm QCD}^2/m_Q^2)$. We use explicit light quark masses $m_u=2.1$ MeV and $m_s = 93$ MeV \cite{PhysRevD.98.030001}.  
We can test this dependency by determining $\tilde{\cal C}_0$ and ${\cal C}_2$ for both the $j_l^P=\frac{1}{2}^-$ and $\frac{3}{2}^+$ multiplets  using the observed masses of the $D_u, D_s$ and $B_u, B_s$ systems \cite{PhysRevD.98.030001}.   The resulting $\tilde{\cal C}_0$ and ${\cal C}_2$ values are shown in Table \ref{tab:lightdyn}, where we see that the  dynamic coefficients have weak dependence on light quark masses between $m_u\to m_s$.
\begin{table}[htbp]
   \centering
   \begin{tabular}{@{} l|rr|rr @{}}
     \hline \hline
      $\tilde {\cal C}(j_l^P )$ & $D_u$ & $D_s$ & $B_u$ & $B_s$ \\ 
      \hline   
      $\tilde {\cal C}_0(\frac{1}{2}^-)$     & $0.0$   & $14$  & $0.0$ & $-1.0$ \\ 
      ${\cal C}_2(\frac{1}{2}^-)/m_Q$ & $141$ & $144$ & $45$ & $49$ \\
      $\tilde {\cal C}_0(\frac{3}{2}^+)$     & $468$ & $498$ & $418$ & $433$ \\
      ${\cal C}_2(\frac{3}{2}^+)/m_Q$ & $45$  & $40$  & $11.3$ & $11.2$  \\[2pt]
   \hline \hline
   \end{tabular}
   \caption{The variation of $\tilde{\cal C}_0$ and ${\cal C}_2$ with light quark mass.
 All entries are in MeV.}
   \label{tab:lightdyn}
\end{table}

Here we are interested in the $D_s$ and $B_s$ systems. Using the known $\mathcal{O}(1/m_Q)$ behavior in heavy light systems, it is possible to extrapolate physical results to the $m_Q \rightarrow  \infty$ limit.  However, as the $j_l^P= \frac{1}{2}^+$ states in the $B_s$ system are not yet observed,  we take the spin-averaged centre-of-gravity (COG) of these states from the LQCD calculation of Lang, Mohler, Prelovsek and Woloshyn \cite{Mohler:Bs0p}, and use general HQET relations to obtain spin-splittings. Other values are taken from the particle data group \cite{PhysRevD.98.030001}. We define $M_G=(3 M_{Q\bar u}^* + M_{Q\bar u})/4$ as the center-of-gravity of the $Q\bar u$ ground state. The $D^0$ has $M_G= 1971.35$ MeV, and the $B^{\pm}$ has $5313.36$ MeV \cite{PhysRevD.98.030001}. $\Delta S$ is the spin-splitting within a spin-multiplet. Then, the heavy quark mass dependence for heavy-light systems is given in Table \ref{tab:HQET}, where the physical mass is $M =$ shift(cog) $+ \Delta S + M_G$.
\begin{table}[htbp]
   \centering
   \begin{tabular}{@{} l|cr|rr|r @{}}
     \hline\hline
      $j_l^P (J^P)$ & \multicolumn{2}{c|}{$D_ s$}  &  \multicolumn{2}{c|}{$B_s$} & $m_Q \rightarrow \infty$ \\ 
                            & ~shift(cog) & $\Delta S$ & ~~ shift(cog) & $\Delta S$ & ~~shift(cog) \\
      \hline   
      $\frac{1}{2}^- (0^-)$ & $104.9$ & $-107.9$ & $89.9 $ & $-36.4$ & $81.1 $ \\ 
      $\frac{1}{2}^- (1^-)$ & $104.9$ & $36.0$   & $89.9$  & $12.1 $ & $81.1$ \\
      $\frac{1}{2}^+ (0^+)$ & $452.7$ & $-106.3$ & $427.0$ & $-36.0$ & $411.8$ \\ 
      $\frac{1}{2}^+ (1^+)$ & $452.7$ & $35.4$   & $427.0$ & $12.0$  & $411.8$ \\ 
      $\frac{3}{2}^+ (1^+)$ & $589.2$ & $-21.3$  & $523.7$ & $-7.0$  & $485.1$ \\ 
      $\frac{3}{2}^+ (2^+)$ & $589.2$ & $12.7$   & $523.7$ & $4.2 $  & $485.1 $ \\[2pt]
   \hline\hline
   \end{tabular}
   \caption{Heavy quark mass dependence for heavy-light systems. Assumptions about the $B_s(\frac{1}{2}^+) $ states are discussed in the text. All masses are in MeV.}
   \label{tab:HQET}
\end{table}

Decays of $j^P_l = \frac{1}{2}^+$ states involving a pseudoscalar, such as $\frac{1}{2}^+\to \frac{1}{2}^- + 0^-$, can happen through S-wave decays, and are expected to have a large branching fraction.  Conversely, decays of $\frac{3}{2}^+$ states involving a pseudoscalar, such as $\frac{3}{2}^+\to \frac{1}{2}^- + 0^-$, can only happen through D-wave\footnote{For finite $m_Q$, the $j_l^P=\frac{3}{2}^+$ ($J^P=1^+$) state will also have a S-wave decay component due to $\mathcal{O}(1/m_Q)$ mixing with the $j_l^P=\frac{1}{2}^+$ ($J^P=1^+$ ) state.}, and are expected to be narrow.  This is the pattern observed for the $D^{0,\pm}$ and $B^{0,\pm}$ excitation spectrum \cite{PhysRevD.98.030001}. However, for the $Q\bar s$ systems with $m_Q \ge m_{\text{charm}}$, Table \ref{tab:HQET} shows that the mass of the $j_l^P = \frac{1}{2}^+$ multiplet is lower than the lowest S-wave threshold for decay into a up/down quark  $j_l^P = \frac{1}{2}^-$ ground state and a kaon, e.g., the $D_s(2317)$ cannot decay to $D K$. Therefore these states are essentially stable against strong isospin preserving decays\footnote{As an exception, for physical pion masses there is a very small rate for the allowed $1^+ \rightarrow 0^- + 2\pi$ transition.}.
The failure of relativistic quark potential models \cite{Godfrey:1985xj,DiPierro:2001dwf, Godfrey:2016nwn} to predict these states being below threshold was surprising, and led to a variety of new theoretical models for these states which are still valid today \cite{Du:2017zvv,Hwang:2004cd,Bardeen:2003kt} . These models can be disentangled by exploring how the states behave as the proximity to the strong decay threshold is varied. This is the subject  of the next sections.

\section{Chiral Symmetry and the Kaon Mass}
\label{sec:chi}

In the light meson sector, the QCD dynamics are entirely different. The spontaneous breaking of chiral symmetry produces light pseudoscalar bosons with mass
\begin{align}
  M^2_{q_1\bar{q_2}}  = B_0(m_{q_1} + m_{q_2})  -  \frac{1}{2}x \ln(\Lambda_{\chi}^2/M^2) + O(x^2) \,,
  \label{eqn:ChiBinding}
\end{align}
where $B_{0} = \Sigma/F^2 $, and $x = M^2/(4\pi F)^2$. Here, $\Sigma = -\langle \bar uu \rangle$ and $\Sigma^{\frac{1}{3}}  = 272(5)$ MeV, where $\Sigma$ and $F$ are evaluated at zero quark mass \cite{Aoki:2019cca}. 
Corrections away from this limit are small for pseudoscalar masses up to $M_K$. For example,  with $F_{\pi} = 92.2(1)$ MeV then $F_{\pi}/F = 1.077$ and $F_{K}/F_{\pi} = 1.191(160)(17)$ \cite{Bazavov:2009bb}.

Quark potential models also fail to capture these dynamics, most notoriously the massless chiral properties. Although, preserving some features may be possible in chiral quark models \cite{Manohar:1983md}. 

With regards to the kaon, its association with spontaneously broken chiral symmetry becomes less and less valid as the strange quark mass increases.  As the strange quark mass exceeds the scale of $\Lambda_{\rm QCD}$, the kaon mass will no longer be well represented by  Eqn.~(\ref{eqn:ChiBinding}).  With this understanding, we can now describe how it is possible to vary the strange quark mass, with small perturbations away from QCD, in order to alter the heavy-light meson distance from the lowest strong decay threshold.   This will allow us to finally understand the physical mechanisms of heavy-light meson states coupled to nearby thresholds in a theoretically simple way.

\section{Isolating Threshold Effects By Varying The Quark Mass}
\label{sec:simple}

Here we show how it is possible to describe the quark mass dependence of particular hadronic decays. We can then use this dependence to smoothly vary the quark mass to push a bound initial state (which lies below threshold) to above the threshold. Further, we can smoothly choose the amount that we want the initial state to be above or below the threshold, making the decay increasingly kinematically allowed or forbidden. The opposite situation also holds, where we can lower a resonance state to be below threshold and turn it into a bound state.

We will focus on the $B_{s0} \to B K$ and $D_{s0} \to DK$ decays as they have properties that make them theoretically simple for a LQCD calculation. This will be discussed in Sec.~\ref{sec:lattice}. In addition, the $D_{s0}\to D_s \pi$ violates isospin symmetry, and $D_{s0}\to D_s \eta$ is expected to be negligible. We will now describe the quark mass dependence of these two decays using HQET and spontaneously broken chiral symmetry.

\begin{figure}[t]
  \centering
  \includegraphics[width=0.49\textwidth]{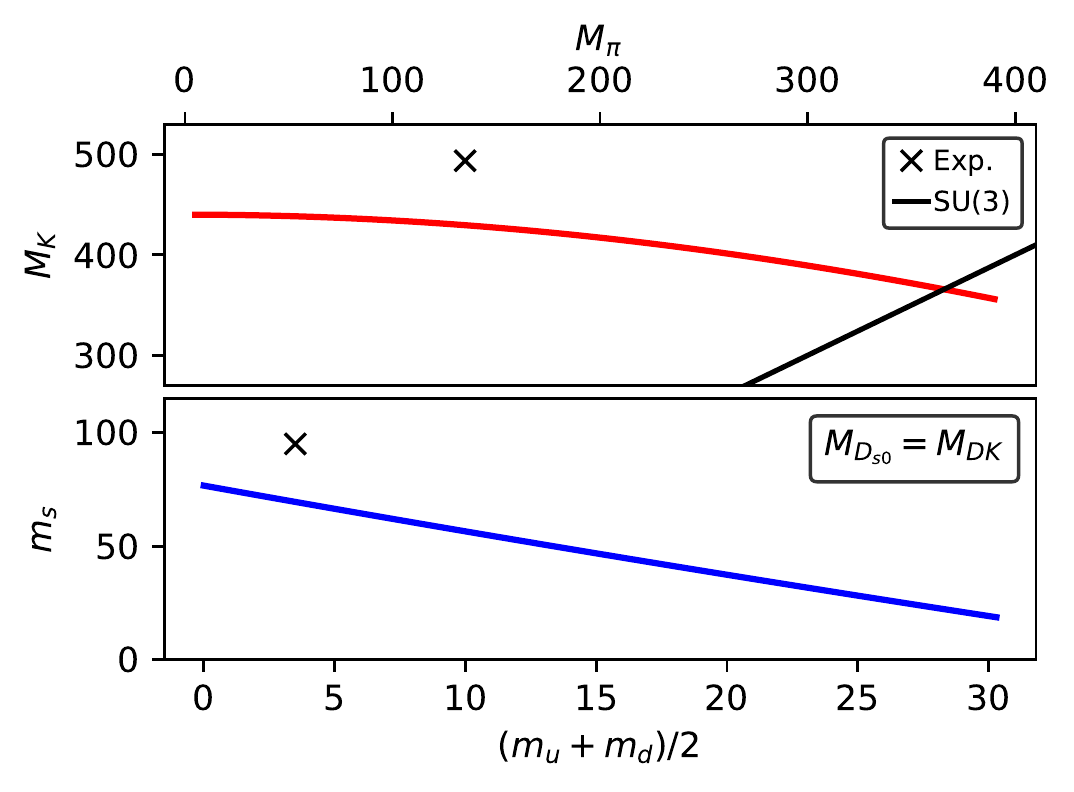}
  \caption{ Color online.  (Upper) The pion and kaon masses needed for the $D_{s0}$ to be exactly at the $D K$ threshold. (Lower) The corresponding values of the strange-quark mass with the up/down-quark averaged mass. The $SU(3)$ symmetric line is when $M_K=M_{\pi}$.  Values derived as discussed in the text.
  }
  \label{fig:Ds0}
\end{figure}

\subsection{$B_{s0} \to B K$ and $D_{s0} \to DK$ }
\label{sec:simpledecay}

We can actively change the value of the strange quark mass in order to explore the effects of this S-wave threshold. Using Eqns.~(\ref{eqn:Hlmass}) and (\ref{eqn:ChiBinding}), consider the quark mass dependence of both the initial $D_{s0}/B_{s0}$ and final $DK/BK$ states in either decay.
The heavy quark dependence is similar in both the initial and final state. However, because the dynamics are different between heavy-light and chiral systems, the strange quark mass dependence of the initial state is different than that of the final state. Explicitly taking the $D_{s0}\to DK$ system as an example, to leading order in chiral perturbation theory, $M_K \propto \sqrt{m_s+m_u}$, but $D_{s0}$ does not have this dependence. 

To highlight how straightforward it is to smoothly make the $D_{s0}/B_{s0}$ lie on the S-wave $DK/BK$ threshold, let $m_s' = m_s - \epsilon$ be the new unphysical strange quark mass. As discussed in Sec.~\ref{sec:HQET}, the binding energy terms ${\cal C}_0$ and ${\cal C}_1$ of Eqn.~(\ref{eqn:Hlmass}) are largely independent of light quark masses ranging from $m_s\to m_u$. In the following we assume the leading order heavy-light mass dependence from HQET, and that the heavy-light binding energy is indeed independent of the small changes in the strange quark mass\footnote{Here we assume that no other strong dynamics alters the simple assumption on quark mass dependence. If additional threshold behavior of the $j_l^P=\frac{1}{2}^+$ states effects binding, this would alter the exact point at which the state is at threshold but not the general conclusion.}. Now, for a particular $\epsilon$, the $D_{s0}/B_{s0}$ mass decreases by an additive shift of $-\epsilon$ MeV. However, using leading order chiral perturbation theory from Eqn.~(\ref{eqn:ChiBinding}) for the kaon mass, $M_K^2$ changes by $-B\epsilon$ MeV$^2$, where $B \sim 2$ GeV (c.~f.~Sec.~\ref{sec:chi}). Consequently, by reducing the strange quark mass, the $D_{s0}/B_{s0}$ mass decreases slower than the $DK/BK$ threshold rest mass. Therefore it is possible to choose a magic value of $m_s'$ where the two masses are identical.

\begin{figure}[t]
  \centering
  \includegraphics[width=0.49\textwidth]{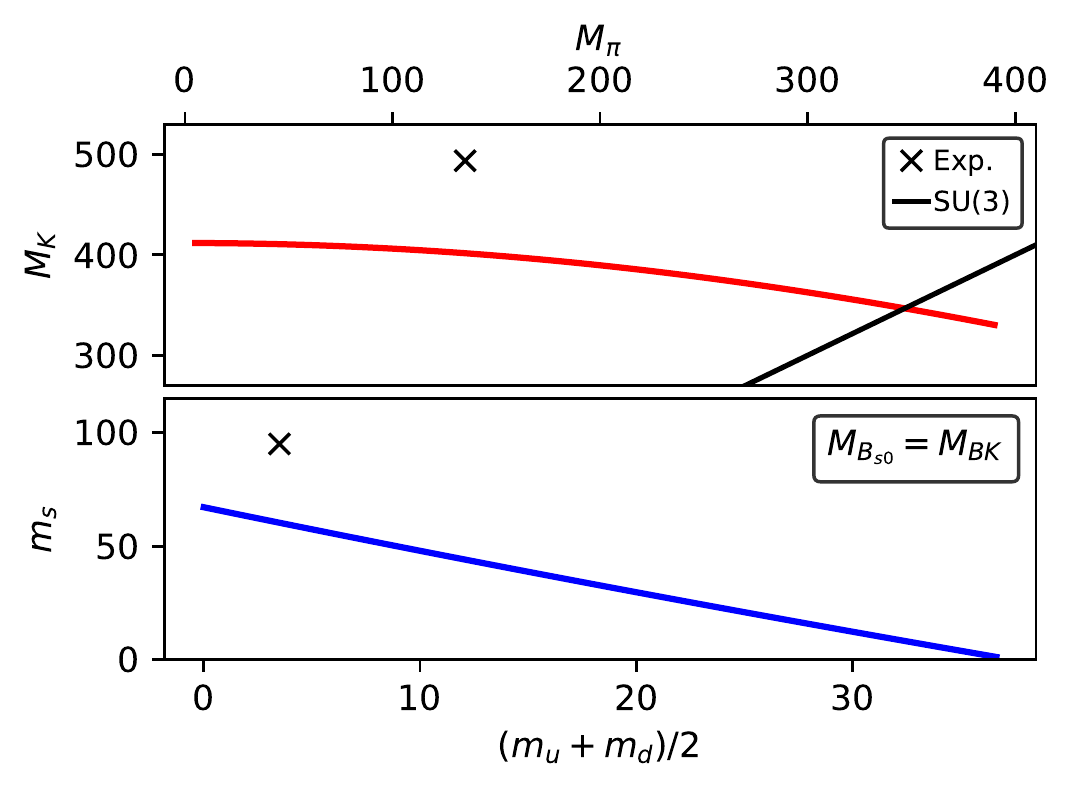}
  \caption{ Color online.  (Upper) The pion and kaon masses needed for the $B_{s0}$ to be exactly at the $BK$ threshold. (Lower) The corresponding values of the strange-quark mass with the up/down-quark averaged mass. The $SU(3)$ symmetric line is when $M_K=M_{\pi}$. Values derived as discussed in the text.
  }
  \label{fig:Bs0}
\end{figure}

Using the methodology just described, we show the magic value of $m_s'$ (where $M_{D_{s0}}=M_{DK}$) as a function of the up/down quark mass in Fig.~\ref{fig:Ds0}. Fig.~\ref{fig:Bs0} shows the magic strange quark mass needed for the $B_{s0}$ system. The experimental values \cite{PhysRevD.98.030001} for $(m_u+m_d)/2$, $m_s$, $M_{\pi}$, $M_K$, $M_{D}$, $M_{B}$, $M_{D_{s0}}$, as well as $M_{B_{s0}}$ from LQCD \cite{Mohler:Bs0p}, were used to determine the magic strange quark values. Notably, by accident, the $D_{s0}/B_{s0}$ states are unique in that they are very close to the lowest threshold, and to move them to this threshold only requires small changes in the strange quark mass. This means that the information on threshold effects obtained from using $m_s'$ in this system can be applied to models in nature which have strange quark mass $m_s. $\footnote{Note that this is not always the case. If there was a large change away from the physical quark masses, then any new mechanisms found in the unphysical theory may not easily apply to the physical theory. In which case, this approach may not be useful to resolve any discrepancies between experiment and models. } If the initial and final state masses were well separated, then changing the strange quark mass may not have made them overlap.

A nice feature of the $D_{s0}/B_{s0}$, in contrast to the $D_{s1}(2536)/B_{s1}(5830)$ and  $D_{s2}^*/B_{s2}^*$ systems,  is that the magic strange quark value is smaller than the physical value.\footnote{The $D_{s1}(2460)/B_{s1}$ states behave exactly like their $D_{s0}/B_{s0}$  ($j_l^P = \frac{1}{2}^+$) partners, hence all the following discussion would be identical for these states.} 
This ensures that the kaon moves closer to the chiral limit, increasingly validating our chiral behavior assumption. Further, choosing a strange quark mass smaller than the magic values shown in Figs.~\ref{fig:Ds0} and \ref{fig:Bs0} would cause the $D_{s0}/B_{s0}$ to be above threshold by an adjustable amount. Studying the $D_{s0}/B_{s0}$ states as they cross through threshold will give additional information which can help understand the S-wave threshold effects of QCD.

We have identified the simplest theoretical system in which varying the strange-quark mass could be used to change the distance of the state from the lowest S-wave threshold. Obtaining information on how the state changes as the threshold is approached would provide useful information that can be used to understand QCD threshold effects.
The calculation that varies the strange-quark mass in the $D_{s0}/B_{s0}$ states can be performed using LQCD. As such, the rest of this paper is concerned with how practical it is to perform this LQCD calculation.

\begin{figure}[t]
  \centering
  \includegraphics[width=0.49\textwidth]{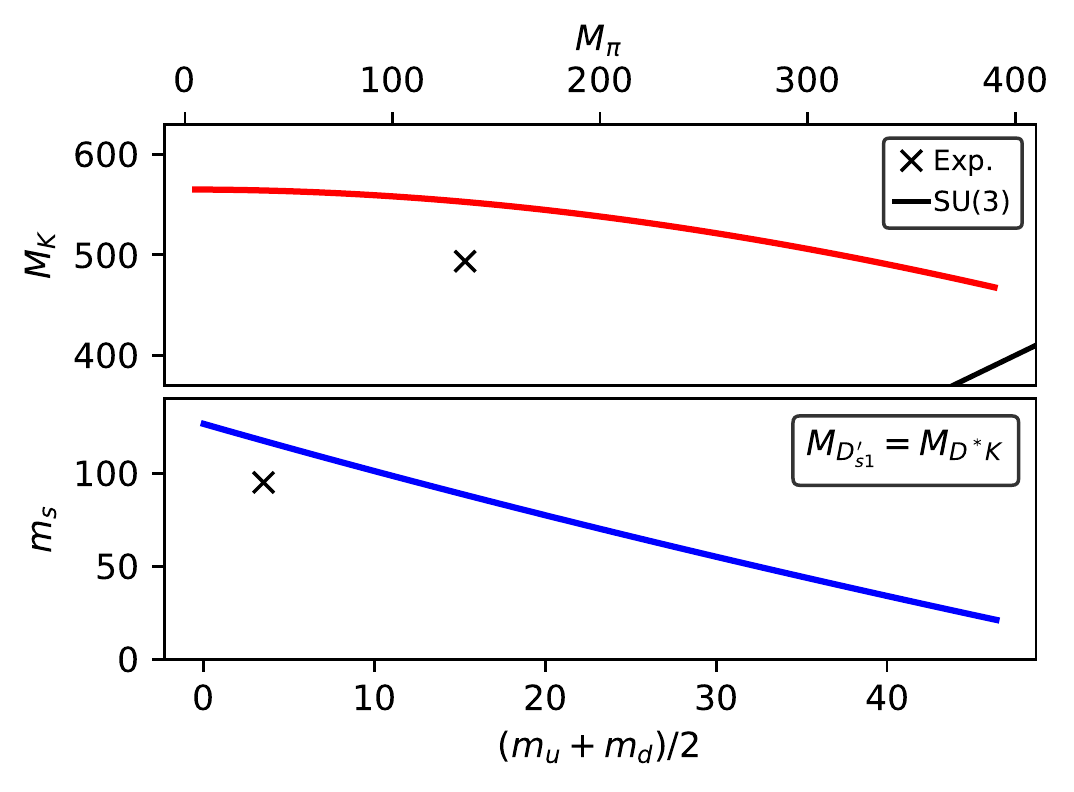}
  \caption{Color online.  (Upper) The pion and kaon masses needed for the $D_{s1}'(2536)$ to be exactly at the $D^*K$ threshold. (Lower) The corresponding values of the strange-quark mass with the up/down-quark averaged mass. The $SU(3)$ symmetric line is when $M_K=M_{\pi}$. Values derived as discussed in the text.
  }
  \label{fig:Ds1}
\end{figure}

\subsection{Other Heavy-Light Decays}

Although the $D_{s0}/B_{s0}$ states are the simplest heavy-light states to study theoretically in order to quantify S-wave threshold effects, additional useful information can be obtained from the $j_l^P = \frac{3}{2}^+$ decays. Experimentally, the $j_l^P = \frac{3}{2}^+$ states, which have $J^P=1^+$ and $2^+$, are narrow and lie above the $D^*K/B^*K$ thresholds. This is in line with quark model predictions \cite{Chen:2016spr}. By the same analysis as done above, the strange quark mass would need to be increased in order to make the $j_l^P = \frac{3}{2}^+$ states lie on top of the relevant threshold. However, raising the strange quark mass could move the kaon out of the chiral regime discussed in Sec.~\ref{sec:chi}. To ensure the smallest change of the strange quark mass is needed, we focus on the $D_{s1}'(2536)/B_{s1}'(5830)$ states since these are closer to the $D^*K/B^*K$ threshold (compared to the $J^P=2^+$ states).

\begin{figure}[t]
  \centering
  \includegraphics[width=0.49\textwidth]{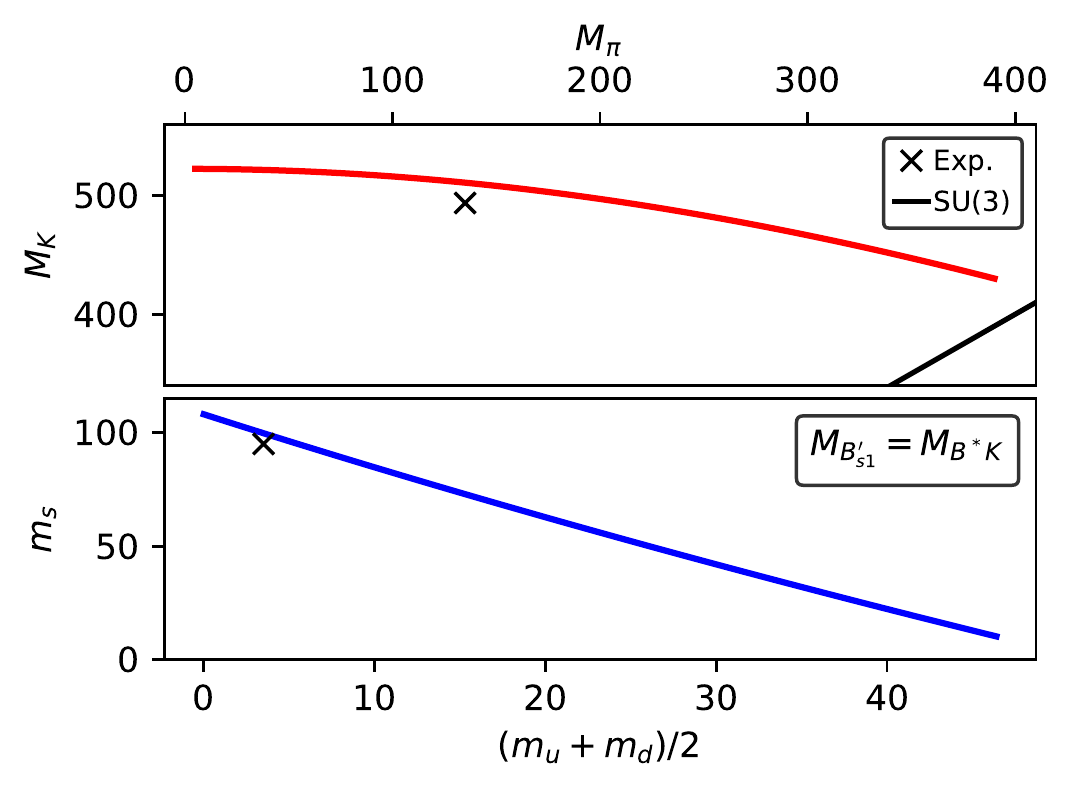}
  \caption{Color online.  (Upper) The pion and kaon masses needed for the $B_{s1}'(5830)$ to be exactly at the $B^*K$ threshold. (Lower) The corresponding values of the strange-quark mass with the up/down-quark averaged mass. The $SU(3)$ symmetric line is when $M_K=M_{\pi}$. Values derived as discussed in the text.
  }
  \label{fig:Bs1}
\end{figure}

These $D_{s1}'/B_{s1}'$ states have two decay modes. One is through a D-wave $D^*K/B^*K$, which is allowed in leading order HQET. The other is through mixing with the $J=1^+$ state which has $j_l^P=\frac{1}{2}^+$, which can decay through a S-wave $D^*K/B^*K$ channel. This process occurs at $\mathcal{O}(1/m_Q)$ in HQET. Both modes are expected to be small. As such, these states would be narrow both above and below threshold. 

This is in contrast to the other P-wave $J=1^+$ state which have $j_l^P=\frac{1}{2}^+$. In leading order HQET these states, if above threshold, can decay through the S-wave $D^*K/B^*K$ mode. Such states are expected to have a wide width. By pushing the $J=1^+$ state with $j_l^P=\frac{3}{2}^+$ below threshold, and comparing to the physical $j_l^P=\frac{1}{2}^+$ state, this would give additional information whether strong S-wave two-meson effects are important for the observed $j_l^P = \frac{1}{2}^+$ states. Consequently, this can distinguish between the various models of the $D_{s0}(2317)$, some of which say two-meson effects are important \cite{Du:2017zvv,Hwang:2004cd}, while others do not \cite{Bardeen:2003kt}. Under the assumption that the chiral behavior still holds for the kaon when the strange quark mass is changed, the magic strange quark masses needed to make the $D_{s1}'(2536)/B_{s1}'(5830)$ reach the S-wave $D^*K/B^*K$ threshold are shown in Figs.~\ref{fig:Ds1} and \ref{fig:Bs1}. These figures were derived using the methodology described in Sec.~\ref{sec:simpledecay}, and with experimental values for $M_{B^*}$, $M_{D^*}$, $M_{D_{s1}'}$ $M_{B_{s1}'}$ \cite{PhysRevD.98.030001}. 
As a last remark, this system may exhibit similar behavior to the $X(3872)$, but is much simpler to study.

\section{Positive Prospects for study in Lattice QCD }
\label{sec:lattice}

As described above, we have found a computationally straightforward methodology that can illuminate the mechanisms behind S-wave threshold effects in QCD. As the $D_{s0}$/$B_{s0}$ states are accidentally very close to the lowest S-wave (and only appreciable) $DK/BK$ threshold in nature, varying the strange-quark mass can push the $D_{s0}$/$B_{s0}$ states closer to the threshold. Our proposal involves smoothly varying the strange-quark mass in a LQCD calculation of either the $D_{s0}$ or $B_{s0}$ mass. 

Before describing how pragmatic this proposal is, it is useful to describe how a typical LQCD spectrum calculation is performed. We point the reader to \cite{Gattringer:2010zz} for more details. Given an interpolating operator $\mathcal{O}_{J^{PC}}$ built from valence quark and/or gluon fields, expectation values of this operator can be found by using the two-point correlation function
\begin{align}
  C_{2pt}({\bf{P_{\text{tot}}}}, t_1-t_0) & = \langle \mathcal{O}_{J^{PC}}^{\dagger}(t_1) \mathcal{O}_{J^{PC}}(t_0) \rangle \label{eqn:correxp} \\
  & = \sum_n |\langle 0 | \mathcal{O}_{J^{PC}} | n \rangle|^2 e^{-M_n (t_1 - t_0)}\,,
  \label{eqn:corrdecomp}
\end{align}
where additional indices have been suppressed for clarity. Here, in the second line the two-point correlator has been spectrally decomposed in the Hilbert space formalism, $|0\rangle$ is the fully interacting vacuum, ${\bf{P_{\text{tot}}}}$ is the total three-momentum of the operator, and $t_1-t_0$ is the propagation time. In principle, the masses $M_n$ of all finite-volume $J^{PC}$ (potentially multi-body) states $|n\rangle$ can be extracted from the multi-exponential decay of this function. 

LQCD calculations numerically evaluate $C_{2pt}$ by replacing the Feynman path-integral of (\ref{eqn:correxp})  with a finite sum over $N$ configurations. As such, one needs an ensemble of configurations, $\mathcal{U}_{m_{q_i},\beta} = \{U_1, U_2,\ldots, U_N \}$, to evaluate the sum. Each $U_j$ is a four-dimensional Euclidean lattice which contains gauge-links - Lie group elements which encode the gauge-fields - sitting on each link between lattice sites. Notably, each $U_j$ is generated with importance sampling according the Boltzmann probability distribution depending on a specific discretized sea-action, namely $S = S_{\text{YM}} + \sum_{i=1}^{{N_f}}S^{\text{sea}}_{q_i}$. Here, $S_{\text{YM}}$ is the gluon Yang-Mills action with gauge coupling encoded in the $\beta$ parameter, and each $S^{\text{sea}}_{q_i}$ is the sea-quark action with mass parameter $m_{q_i}$.
Each of these mass parameters do not have to be set to their physical values, and can be taken to be unphysical if required. Typically this is done to make calculations less expensive or to guide extrapolations. ${N_f}$ is the number of sea-quarks, and usually $N_f = 2 + 1 + 1$, meaning there are $2$ light-quarks of equal mass (conceptually the equal mass up- and down-quarks), the strange-quark, and the charm quark. The bottom- and top-quarks are not included in the sea, and such effects are taken to be negligible.

In LQCD calculations involving quarks, a valence quark propagator is needed for each configuration of the ensemble. The valence quark propagator on each configuration is found as the Green's function to the valence quark action $S^{\text{val}}_{q_i}[U_j]$ kernel. Notably, the discretized valence-quark action does not need to match the sea-quark counterpart. In principle, even the valence- and sea-quark mass value do not need to match. This scenario is called partially-quenching, and may lead to unitarity violations. However, as generating the sea-quarks in ensembles can be one of the most expensive parts of a LQCD calculation, using multiple different valence- quark masses not tuned to their sea counterpart can be common in order to explore as much physics as possible for the smallest cost \cite{Fermilab:ensembles}.

Having already discussed the positive impact of studying the $D_{s0}$/$B_{s0}$ states, which can chosen to be either above or below threshold, we now discuss how straightforward the LQCD calculation is, and demonstrate that it is an extension to works already present in the literature.

\paragraph{Finite-Volume Effects Below and Above Threshold.}

All LQCD calculations are performed in a four-dimensional Euclidean box with temporal extent $T$ and spatial extent $L$. Ensuring that the box corresponds to zero temperature requires $T>L$. Finite-volume effects then need to be quantified. For single-particle bound states, L\"{u}scher has shown that finite-volume effects from ``around-the-world interactions'' are exponentially suppressed with $M_{\pi}L$ \cite{Luscher:1985dn}. Conventional knowledge takes $M_{\pi}L\ge 4$ as sufficient to neglect these finite-volume effects. 

However, the situation is dramatically different for states above threshold. L\"{u}scher has also shown that the finite-volume corrections to the non-interacting two-hadron mass, which can be calculated in LQCD, can be used to extract the Minkowski space pole of the scattering matrix \cite{Luscher:1986pf}. As such, when the $D_{s0}$/$B_{s0}$ state is below threshold and is bound, the LQCD calculation just needs to be done on a single volume, as in \cite{Mohler:Ds0p,Mohler:Bs0p}. Still, the volume needs to chosen large enough to yield the correct virtual two-meson contributions, as shown in Fig.~$10$ of \cite{Bali:Ds0p}, where a binding of $\sim 30$ MeV requires a box with length $L=4$ fm.

However, when the $D_{s0}$/$B_{s0}$ state is above threshold and is a resonance, the L\"{u}scher method needs to be employed. As we are only interested in states close to the S-wave threshold, the phase shift can be expanded in the effective range approximation \cite{Mohler:Dpi}. On a single ensemble, fitting this functional dependence only requires two energy levels: the finite-volume energy levels associated with the threshold and the would-be resonance \cite{Mohler:Ds0p}.

In both situations, although not necessary, more information would help determine the $D_{s0}$/$B_{s0}$ state pole mass more precisely. For example, more data could be obtained on the same ensemble by using more operators which have been subduced into lattice irreducible representations \cite{Dudek:2012xn, Mohler:Dpi}, or by using another ensemble with a different volume but with other scales kept fixed \cite{Bali:Ds0p}.  

Consequently, a significant number of ensembles that already exist and are used by the lattice community \cite{Fermilab:ensembles, Mohler:Dpi, Mohler:Ds0p, Bali:Ds0p, Dudek:2012xn, Wilson:2019wfr, Moir:2016srx} can be used for our proposed S-wave threshold study. This is evidenced by the fact that timely LQCD studies have verified the existence of the $D_{s0}$ \cite{Mohler:Ds0p,Bali:Ds0p} and $B_{s0}$ \cite{Mohler:Bs0p} states below threshold. Our proposal is a straightforward extension of those works, but would be significantly impactful in understanding why the state is bound. 

\paragraph{Lattice Spacing.}
With regards to the lattice spacing, most modern calculations have lattice spacings $a < 0.1$ fm. This is sufficiently small so that finite lattice spacing effects are unlikely to change any S-wave threshold effects, or the mechanism for binding. Since our goal is to understand the S-wave threshold effects, it is more useful to perform three different strange-quark masses at one lattice spacing, rather than three different lattice spacings at one strange quark mass. As such, a single lattice spacing can be used throughout and a continuum extrapolation is not necessary. If necessary, ratios of hadron masses can be used to help remove systematic errors from lattice spacing corrections. 

\paragraph{Signal-to-Noise.}
An important consideration for LQCD studies is how quickly the statistical errors become excessively large, prohibiting the ability to practically extract useful information. The conventional Lepage-Parisi argument \cite{Lepage:1989hd, PARISI1984203} says that the noise in the expectation value of an operator is controlled by the square root of the variance of that operator. From Eq.~(\ref{eqn:corrdecomp}), we can see that the variance will be set by the lowest state which contributes to $\langle (\mathcal{O}^{\dagger}_{J^{PC}}\mathcal{O}_{J^{PC}})^{\dagger} (\mathcal{O}^{\dagger}_{J^{PC}}\mathcal{O}_{J^{PC}})\rangle$.

The systems we need to study are the $D_{s0}$/$B_{s0}$, the $D/B$ and the $K$. Note that the $K/D/B$ meson masses are only needed if the L\"{u}scher method is being used, as this requires the non-interacting threshold mass on each ensemble as input, e.g., $M_K + M_D$. The finite-volume $DK/BK$ rest mass is close to the $D_{s0}$/$B_{s0}$ mass by construction, and so similar signal-to-noise arguments apply to both. For the $D_{s0}$/$B_{s0}$, the lowest state in the variance is the $\eta_c+\eta_s/\eta_b+\eta_s$,\footnote{The $\eta_s$ is a stable state from LQCD calculations of $\bar s s$ pseudoscalar mesons where the strange quarks are not allowed to annihilate. This $\eta_s$ particle has a mass $M_{\eta_s}$ of $689$ MeV \cite{Dowdall:2011wh}. LQCD determinations of $M^2_{\eta_s}$ agree with the leading order chiral perturbation theory $2M^2_K - M^2_{\pi}$ value to within $1\%$ \cite{Dowdall:2011wh}.  } and so for the $D_{s0}$ the signal-to-noise at large time behaves as $\sim\exp(-(M_{D_{s0}} - (M_{\eta_c} + M_{\eta_s})/2) t)$, and similarly for the $B_{s0}$. This mass splitting is around $480/670$ MeV for the $D_{s0}/B_{s0}$ \cite{PhysRevD.98.030001, Mohler:Bs0p}, which is well within the bounds of accurate LQCD calculations. For example, see Fig.~$4$ of \cite{Bali:Ds0p} to examine the signal-to-noise for the physical $D_{s}(2317)$. 

For the $D/B$ mesons, the lowest state in the variance is the $\eta_c+\pi/\eta_b+\pi$ state, and so the signal-to-noise behaves as $\sim\exp(-(M_B - (M_{\eta_b} + M_{\pi})/2 )t)$. This mass splitting is around $500$ MeV, which is not prohibitive and there are many precision physics calculations of $B$-mesons in the literature \cite{Bazavov:2019aom, Fermilab:ensembles, Davies:2019gnp, Hughes:2017spc, Colquhoun:2015oha}. For the $D$-meson, the signal-to-noise mass splitting is $300$ MeV. Similar arguments can also be applied to the $K$ to show that it has virtually no signal-to-noise problem. Consequently, the lattice data will be sufficiently accurate to extract a good determination of both the finite-volume $D_{s0}/B_{s0}$ mass and the S-wave $DK/BK$ rest mass. 

As mentioned, we propose to vary the strange-quark mass by a small amount to determine how the mass of the $D_{s0}$/$B_{s0}$ changes. These variations of the strange-quark mass will not change the above arguments appreciably. As such, signal-to-noise issues should not prohibit this proposal. 

\paragraph{Position of the States in the Spectrum.}
With rotational symmetry, any interpolating operator $\mathcal{O}_{J^{PC}}$ will create all states which have the same $J^{PC}$ quantum numbers. As such, one needs to extract the mass of the state of interest, $|n'\rangle$, from the multi-exponential decomposition in Eq.~(\ref{eqn:corrdecomp}). Non-ground state contributions decay away exponentially fast, and if the signal-to-noise also decays exponentially fast, then extracting non-ground state observables becomes computationally difficult\footnote{Although not necessary for this proposal, the identification of states high in the spectrum is possible by using a large array of operators and the variational method \cite{Peardon:2009gh, Dudek:2012xn}.}.

In our proposal however, all states are the lowest in the spectrum. Taking the $B_{s0}\to B K$ as an example, the $B$ and $K$ are the lowest states in their respective $0^-$ channels, and the $B_{s0}$ is the ground state of the $0^+$ channel.

Still, it is important to consider that the $D_{s0}/B_{s0}$ state is close to the $DK/BK$ threshold in nature \cite{PhysRevD.98.030001, Mohler:Bs0p}, with a binding energy of around $ 30$ MeV. Being so close to threshold makes the extraction of the $D_{s0}/B_{s0}$ mass slightly more difficult. In a LQCD calculation, this can be seen from the spectral decomposition in Eq.~(\ref{eqn:corrdecomp}), where two nearby exponentially decaying contributions can be difficult to separate.

If $M_{B_{s0}}$ and $M_{BK}$ are sufficiently close, then it can be difficult to separate the two exponential contributions \cite{Dudek:2012xn} when only using meson interpolating operators that look like the single particle $B_{s0}$. In this case, only one (incorrect) mass is extracted, which corresponds to some incorrect combination of the two nearby masses, e.g., see Fig.~$6$ of \cite{Bali:Ds0p}. To project out the two correct contributions, it is necessary to include both single meson $B_{s0}$ and two-meson $BK$ interpolating operators. While needing two-meson interpolating operators at rest costs more computational resources, it is by no means prohibitive, evidenced by the multitude of LQCD calculations that utilise two-meson operators \cite{Mohler:Ds0p,Mohler:Bs0p,Bali:Ds0p,Dudek:2012xn,Hughes:2017xie}.

It should also be noted that our proposal is to vary the strange quark mass to bring the $D_{s0}/B_{s0}$ mass as arbitrarily close, and through, the S-wave $DK/BK$ threshold. Practically however, at some point the masses of the state and threshold will be indistinguishable within the statistical error, and this region should be avoided.

\paragraph{Disconnected diagrams.}

As described above, since both the single-meson and two-meson states are close to each other in the spectrum, both types of interpolating operators need to be used. The single-meson operators are straightforward, and have no disconnected diagrams. However, the $DK/BK$ interpolating operators have Wick contractions that require evaluation of light-quark disconnected/annhilation contributions. These can be computationally expensive. Standard approaches to evaluate these contributions are the sequential stochastic \cite{Bali:Ds0p} or the distillation methodology \cite{Peardon:2009gh, Mohler:Ds0p, Mohler:Bs0p}. These have been used to study the $D_{s0}/B_{s0}$ already \cite{Mohler:Ds0p, Mohler:Bs0p, Bali:Ds0p}. 

It should be mentioned that in our proposal the disconnected contributions only need to be evaluated once ever. There is only a single strange-quark in the two-meson operator, and so it is {\it{not}} necessary to recompute the disconnected light-quark components after each change of the strange-quark mass/propagator. 

\paragraph{Mixing With Other Channels.}

The reason we choose the $D_{s0}/B_{s0}$ system is its simplicity. LQCD calculations have shown the existence of the bound $D_{s0}(2317)$ and $B_{s0}$ states when only the elastic S-wave threshold is taken into account \cite{Mohler:Ds0p, Mohler:Bs0p, Bali:Ds0p}. No other channels are needed. As we want to illuminate how the elastic S-wave threshold interacts with mesons, no other channels need to be considered.

\paragraph{Fixed Sea-quarks in Ensembles and Partially-Quenching.}

Most lattice ensembles have $N_F = 2+1$ or $N_F = 2+1+1$ flavors in the sea. As such, isospin will be an exact symmetry in these LQCD calculations. As computational resources grow with smaller light-quark masses, there exists ensembles that have light-quarks which produce pion masses ranging from $\sim 130-350$ MeV. Any of these ensembles are suitable for our proposal, although ensembles closer to the physical mass are more favorable so that the interplay between chiral effects and confinement are correct. 

It is necessary to have the same sea and valence light-quark mass in the LQCD calculation of the threshold $DK/BK$. If not, this would cause appreciable distortions of the correct finite-volume two-meson mass, where an accurate value is needed to project out the $D_{s0}/B_{s0}$ state from the correlator as mentioned above. Because of this, varying the light-quark mass would require entirely new ensembles to be generated, which is prohibitively expensive. 

Instead, we propose to fix the light- and strange-quarks in the sea, but smoothly vary the valence strange-quark mass. In a LQCD calculation, this only requires the re-calculation of the valence strange-quark propagators, which are the one of the numerically inexpensive parts. Having a sea strange-quark that differs from the valence counterpart is conventionally known as partially quenching. Partially quenching has been utilised extensively in LQCD \cite{Fermilab:ensembles}. As we only change the valence strange-quark mass by perturbations around the physical point, such partially quenching effects should be small. Based on Figs.~\ref{fig:Ds0} and \ref{fig:Bs0}, at physical pion mass, to make the $D_{s0}/B_{s0}$ state sit at threshold requires a $15\%$ downward shift of the Kaon mass. This in turn translates into a $30\%$ downward shift of the strange-quark mass. The consequences of this partially quenching shift should be to slightly change the running of $\alpha_s$. 
 
\par

This LQCD project proposal is to quantify S-wave threshold effects by varying the valence strange-quark mass until the $D_{s0}/B_{s0}$ state passes through the $DK/BK$ threshold and becomes a resonance state. After accounting for the various LQCD constraints above, there are no prohibitive issues, and this impactful calculation could be performed on current hardware by multiple collaborations. 

\section{Summary and Discussion}
\label{sec:conclusions}

In this work we have shown how it is possible to describe the quark mass dependence of particular hadronic decays.
We focus on the $D_{s0}\to DK$ and $B_{s0} \to BK$ channels because they are the cleanest theoretically.
In Sec.~\ref{sec:simpledecay} we show how small changes of the strange quark mass can move the $D_{s0}/B_{s0}$ mass to lie on top of the $DK/BK$ threshold.
We do this by using heavy-quark effective theory in Sec.~\ref{sec:HQET} to describe heavy-light states, and chiral perturbation theory in Sec.~\ref{sec:chi} to describe the kaon. The magic strange quark mass where the states lie on the thresholds are shown in Fig.~\ref{fig:Ds0} and \ref{fig:Bs0}.

This lattice QCD project proposal is to quantify S-wave threshold effects by smoothly varying the valence strange-quark mass, at a fixed pion mass near the physical point, until the $D_{s0}/B_{s0}$ state passes through the threshold and becomes a resonance.
In Sec.~\ref{sec:lattice}, we describe how such calculations are practical, the properties of these states that make them theoretically clean to study in lattice QCD, and how this study is a straightforward extension of work already present in the literature (requiring minimal extra computational resources). 
The choice between studying either the $D_{s0}$ or $B_{s0}$ depends solely on which is easier to implement in an existing LQCD codebase. 
Notably, once a single calculation of the $D_{s0}/B_{s0}$ has been performed as in \cite{Bali:Ds0p, Mohler:Ds0p, Moir:2016srx}, our proposal entails: (i) re-computing the strange-quark propagator with a slightly different valence-quark mass; (ii) re-using the old light-/charm-quark propagators with the new strange-quark propagator to compute two-point correlators for $D_{s0}/B_{s0}$ and the two-meson S-wave $DK/BK$ at rest; (iii) fit this data to extract the finite-volume single-particle mass and two-meson rest mass; (iv) if necessary, use the effective range approximation \cite{Mohler:Ds0p, Bali:Ds0p} with the L\"{u}scher method to extract $M_{D_{s0}}/M_{B_{s0}}$; (v) plot $M_{D_{s0}}-M_{D}-M_{K}$ or $M_{B_{s0}}-M_{B} - M_K$ vs.~ $M_{K}$. Additionally, the composition of these states as a potential mixture of single and/or two-meson states could be determined \cite{Bali:Ds0p}.
Multiple existing collaborations could perform this calculation. 

Given the large literature on the XYZ states \cite{Brambilla:2010cs}, an appreciable amount of resources are being spent trying to understand the effects of QCD thresholds. However, a complication in doing so theoretically is including the multiple competing physical processes. This makes theoretical predictions difficult \cite{Cheung:2017tnt}. Here we propose to supplement the experimental data with lattice QCD by varying the strange quark mass and understanding S-wave threshold effects in the cleanest QCD system, which only has one channel. Our understanding of this one system can then help build models of the more complicated scenarios. This in turn could finally give insight into the phenomenology of meson states located near thresholds. In the short term we could explain the nature of the P-wave heavy-light $D_{s0}(2317)$ state below threshold, and in the long term could make progress towards resolving the suspected four-quark dynamics within the XYZ states. 
\newline
\section*{ACKNOWLEDGMENTS}

This manuscript has been authored by Fermi Research Alliance, LLC under Contract No.~DE-AC02-07CH11359 with the U.~S.~Department of Energy, Office of Science, Office of High Energy Physics. 
EE acknowledges the support of the Alexander von Humboldt Foundation  and the hospitality of the Munich Institute for Astro- and Particle Physics (MIAPP) of the DFG cluster of excellence  ``Origin and Structure of the Universe" and also the Institute for Advanced Study at TUM where his research was performed.

\appendix

\bibliographystyle{apsrev4-1}
\bibliography{threshold}

\end{document}